% ...

\documentclass{emulateapj}

\usepackage{color}
\usepackage{graphicx}
\usepackage{epsfig}
\usepackage{epstopdf}
\usepackage{natbib}
\newcommand{\degree}{\ensuremath{^\circ}}
\newcommand{\gamDor}{$\gamma$ Dor}
\newcommand{\Herschel}{\textit{Herschel}}
\newcommand{\Spitzer}{\textit{Spitzer}}
\newcommand{\IRAS}{\textit{IRAS}}

\newcommand{\p}{$\pm$~}
\newcommand{\bout}{$\sim$}
\newcommand{\logg}{$\log (g)$}
\newcommand{\Teff}{$T_{\rm{eff}}$}

\newcommand{\ch}{\colhead}

\newcommand{\tnm}{\tablenotemark}
\newcommand{\tnt}{\tablenotetext}
\newcommand{\chisq}{$\chi^2$}

\definecolor{grey}{RGB}{132,132,132}
\definecolor{green}{RGB}{0,201,87}

\shorttitle{\gamDor's debris disk}
\shortauthors{Broekhoven-Fiene et al.}

\begin{document}

\title{The debris disk around $\gamma$ Doradus resolved with Herschel}

\author{Hannah Broekhoven-Fiene\altaffilmark{1,2}, 
Brenda~C.~Matthews\altaffilmark{2,1}, 
Grant M.~Kennedy\altaffilmark{3}, 
Mark Booth\altaffilmark{1,2}, 
Bruce Sibthorpe\altaffilmark{4}, 
Samantha M.~Lawler\altaffilmark{5}, 
JJ Kavelaars\altaffilmark{1,2}, 
Mark C.~Wyatt\altaffilmark{3}, 
Chenruo Qi\altaffilmark{5,2}, 
Alice Koning\altaffilmark{1,2}, 
Kate Y. L. Su\altaffilmark{6},
George H. Rieke\altaffilmark{6},
David~J.~Wilner\altaffilmark{7}, 
and Jane~S.~Greaves\altaffilmark{8}}

\altaffiltext{1}{Department of Physics and Astronomy, University of Victoria, Victoria, BC V8W 3P6, Canada}
\altaffiltext{2}{Herzberg Institute of Astrophysics, National Research Council of Canada, Victoria, BC V9E 2E7, Canada}
\altaffiltext{3}{Institute of Astronomy, University of Cambridge, Madingley Road, Cambridge CB3 0HA, UK}
\altaffiltext{4}{UK Astronomy Technology Center, Royal Observatory, Blackford Hill, Edinburgh EH9 3HJ, UK}
\altaffiltext{5}{Department of Physics and Astronomy, University of British Columbia, 6224 Agricultural Road, Vancouver, BC V6T 1Z1, Canada}
\altaffiltext{6}{Steward Observatory, University of Arizona, 933 North Cherry Avenue, Tucson, AZ 85721, USA}
\altaffiltext{7}{Harvard-Smithsonian Center for Astrophysics, 60 Garden Street, Cambridge, MA 02138, USA}
\altaffiltext{8}{School of Physics and Astronomy, University of St. Andrews, North Haugh, St Andrews, Fife KY16 9SS, UK}

\begin{abstract}

We present observations of the debris disk around \gamDor adus, an F1V star, from the \Herschel~Key Programme DEBRIS (Disc Emission via Bias-free Reconnaissance in the Infrared/Submillimetre). 
The disk is well-resolved at 70, 100 and 160~\micron, resolved along its major axis at 250~\micron, detected but not resolved at 350~\micron, and confused with a background source at 500~\micron.
It is one of our best resolved targets and we find it to have a radially broad dust distribution. The modeling of the resolved images cannot distinguish between two configurations: an arrangement of a warm inner ring at several~AU (best-fit 4~AU) and a cool outer belt extending from \bout55 to 400~AU or an arrangement of two cool, narrow rings at \bout 70~AU and \bout 190~AU.
This suggests that any configuration between these two is also possible.
Both models have a total fractional luminosity of \bout$ 10^{-5}$ and are consistent with the disk being aligned with the stellar equator. 
The inner edge of either possible configuration suggests that the most likely region to find planets in this system would be within \bout55~AU of the star. A transient event is not needed to explain the warm dust's fractional luminosity.  

\end{abstract}



\section{Introduction}
\label{sec:intro}

Debris disks were first discovered when observations with the InfraRed Astronomical Satellite (\IRAS) revealed that Vega, $\beta$ Pictoris, and Fomalhaut were unexpectedly bright at infrared (IR) wavelengths \citep{Aumann1984}. Many main-sequence stars have since been observed to possess IR emission above the expected photospheric level, which is usually attributed to the thermal emission of dust that is heated by the host star(s). The dust is second generation, e.g., produced by ongoing collisions of larger bodies, since the dust's lifetime in orbit is too short for it to be primordial in origin, e.g., originating in the protoplanetary disk \citep{Backman1993}. Models suggest that parent planetesimals of at least 10--100 km feed the dust through destructive collisions, though how objects are initially ``stirred" to high enough collision velocities is unclear (e.g., \citealt{Wyatt2007}). One possibility is the formation of bodies large enough ($>$1000km) to stir planetesimals through dynamical interactions \citep{KenyonBromley2004a,MustillWyatt2009}.

Modelling of the stellar spectral energy distribution (SED) at optical wavelengths allows a comparison with the observed IR flux, and possible detection of an IR excess. Modelling the SED of the excess itself provides a measure of the dust temperature and therefore its location (by making assumptions about the emissive properties of the dust). This method has proven instrumental in relating the properties of debris disks to each other. It is generally sufficient to assume that the dust particles behave like blackbodies, but this approach does not yield the exact disk location because dust at different stellocentric distances can have the same temperature (grain size and dust location are degenerate). 
SED modeling can also yield incorrect dust properties such as lower grain size (compare HD107146, \citealt{Roccatagliata2009} versus \citealt{Ardila2004} and \citealt{Ertel2011}).

Thus resolving a disk allows for a deeper understanding of the system since its configuration is directly observed. Disk sizes determined from resolved images are often found to be $\sim$2--5 times larger than those suggested by the SED with blackbody assumptions (\citealt{Schneider2006,Wyatt2008,Matthews2010,RodriguezZuckerman2012}; Booth et 
al., in press). 
Resolving the disk at more than one wavelength is even more advantageous, as it can reveal whether the observed configuration of the system is wavelength dependent, as it is for Vega \citep{Su2005,Sibthorpe2010},  $\beta$ Leo \citep{Matthews2010,Churcher2011} and others. For example, a disk may appear larger at longer wavelengths if it has two components, and the cooler one (i.e.,  further from the star) dominates at the longer wavelength, as is the case for $\eta$ Corvi \citep{Wyatt2005etaCorvi,Matthews2010}.

As more data are collected on debris disks from observatories with improved sensitivity and resolution, we are moving away from the simple disk models of single narrow dust rings and discovering more complicated shapes and configurations. Many debris disks have been found to contain multiple dust components and/or have extended dust distributions with a large range in behaviour (e.g., HD 107146: \citealt{Ertel2011}, $\beta$ Leo: \citealt{Stock2010}, $\zeta$ Lep: \citealt{Moerchen2007}). Some hosts that have multi-component disks are also planet hosts \citep[e.g., $\epsilon$ Eridani: ][]{Backman2009,Reidemeister2011}. HR 8799 even plausibly shows the existence of a planetary system with warm and cold dust that has planets in the gap \citep{Su2009}. This highlights the rich diversity of observed planetary systems.

Since debris disks were discovered with \IRAS, significant advancement of the field has been made with other observatories such as the \textit{Spitzer Space Telescope} (\Spitzer; \citep{Rieke2005,Beichman2006,Bryden2009}, the James Clerk Maxwell Telescope \citep{Holland1998,Greaves1998}, and now the \textit{Herschel Space Observatory} (\Herschel) by observing the thermal emission of the dust.
(Optical observations of star light scattered by dust are also used to study debris disks.) 
 \Herschel~is well suited for debris disk studies as dust emission is well contrasted against stellar emission at its wavelength range of 70--500~\micron. It is sensitive enough to detect the photospheres of nearby stars (and therefore can better determine whether an excess is present), and its resolution of 6.7\arcsec~at 100~\micron~can probe the sizes of nearby disks. We present observations of \gamDor adus (\gamDor)~and its debris disk that were taken with \Herschel~as part of the DEBRIS (Disc Emission via Bias-free Reconnaissance in the Infrared/Submillimetre) Key Programme \citep{Matthews2010}. Stars observed by DEBRIS are taken from the UNS (Unbiased Nearby Stars) sample \citep{Phillips2010}, a volume-limited sample, and so unbiased toward spectral type, binarity, metallicity, and presence of known planets. 

\begin{deluxetable}{lcccl}
\tablewidth{2.5in}
\tablecaption{\label{tbl:gamdor}Stellar Information for \gamDor}
\tablehead{
\ch{Parameter} 	&& \ch{Value} 		&& \ch{Reference} } 
\startdata
Spectral type 	&& F1 V 				&& \cite{GrayCorbally2006} \\
R.A. (J2000) 		&& ~04:16:01.586 	&& \cite{Hoeg2000} \\
Decl. (J2000) 		&& --51:29:11.933 	&& \cite{Hoeg2000} \\
PM-R.A. (mas yr$^{-1}$) 	&& 101.5 			&& \cite{Hoeg2000} \\
PM-decl. (mas yr$^{-1}$) 	&& 184.7 			&& \cite{Hoeg2000} \\
$V$ magnitude		&& 4.25				&& \cite{Balona1994} \\
Distance (pc) 	&& 20.46 \p 0.15 	&& \cite{Phillips2010}\tnm{a} \\
Age (Gyr)		&& 0.4\tnm{b}		&& \cite{Chen2006} \\
Age (Gyr)		&& 0.82--2.19\tnm{c}	&& \cite{Vican2012}
\enddata
\tnt{a}{Incorporates parallaxes from \cite{vanLeeuwen2007} and \cite{vanAltena1995}.}
\tnt{b}{Estimated uncertainty is a factor of two (0.2--0.8 Gyr) using \cite{Schaller1992} isochrones.}
\tnt{c}{Estimates of ages determined from chromospheric activity and X-ray emission, respectively.}
\end{deluxetable}

The debris disk around \gamDor~(HD 27290, HIP 19893), whose basic parameters are listed in Table~\ref{tbl:gamdor},~was discovered with \IRAS~\citep{Aumann1985,Rhee2007}. It has been detected with \Spitzer~but was not resolved~\citep{Chen2006,Koerner2010}. Here we present images of the \gamDor~disk taken with \Herschel~that are the first to directly constrain the location of the dust. \gamDor~is not known to host any extra-solar planets but is a target for the exoplanet search using Near-Infrared Coronagraphic Imager (NICI) on the Gemini-South 8.1 m telescope \citep{Chun2008,Liu2010}, the results of which are not yet publicly available. In Section~\ref{sec:obs}, we present the \Herschel~observations and ancillary data. We measure the \Herschel~fluxes in Section~\ref{sec:fluxes}. In Section~\ref{sec:results}, \gamDor's photosphere is modeled and used to determine the excesses that are observed in the IR and submillimetre (submm). We present a basic analysis of the images in Section~\ref{sec:basicimanal} and the more detailed modeling of the SED and resolved images in Section~\ref{sec:modeling}. The results are discussed in Section~\ref{sec:discussion} and we summarize our conclusions in Section~\ref{sec:conclusions}.

\section{\textit{Herschel} Observations}
\label{sec:obs}

\begin{figure*}
\includegraphics[trim=2cm 7.25cm 1.75cm 3cm, clip=true, width=5 in]{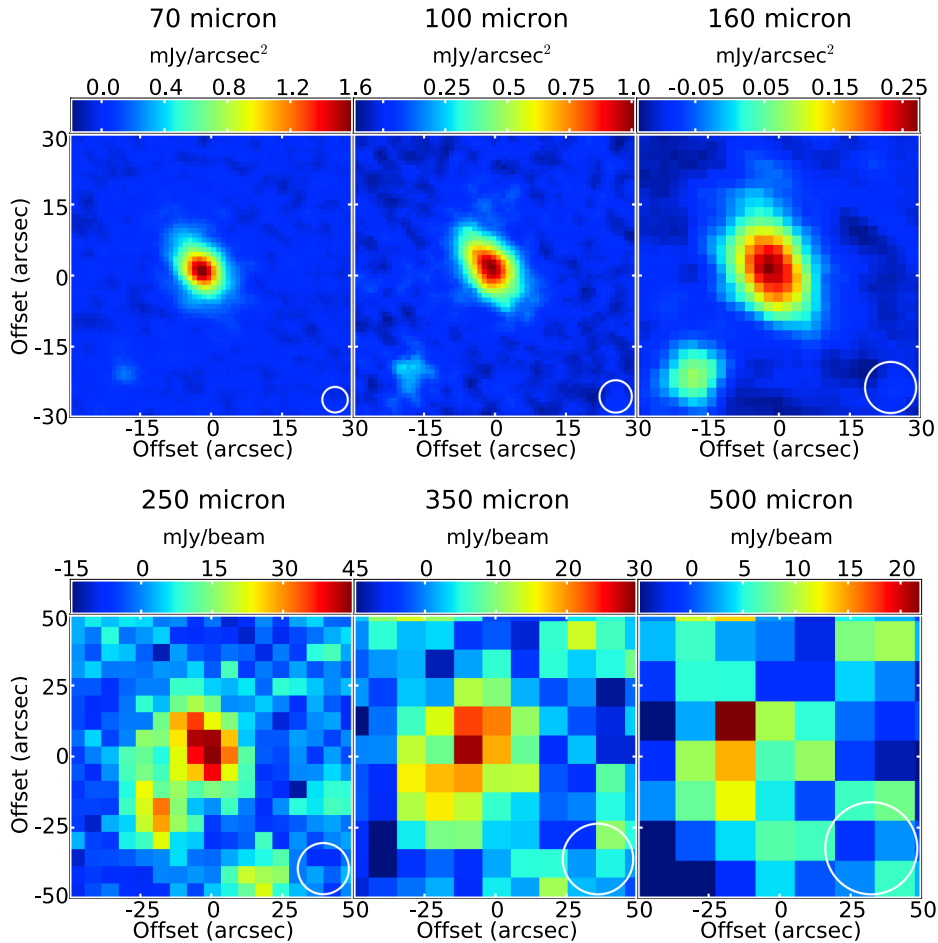}\centering
\caption{PACS (top row) and SPIRE (bottom row) \Herschel~observations showing the 60\arcsec\,$\times$\,60\arcsec~and 100\arcsec\,$\times$\,100\arcsec~regions around \gamDor, respectively. The beam sizes (the FWHMs of the PSFs) are shown in the bottom right corners. A point-like background source is evident \bout30\arcsec~to the southeast of \gamDor~(up is North). The extended emission from \gamDor~is evident in comparison to the background source and the beam sizes. The background source is well separated from \gamDor~at PACS wavelengths, but becomes increasingly difficult to distinguish from \gamDor~at SPIRE wavelengths. At 500~\micron, the flux from both sources is contained within a single beam.}
\label{fig:maps}
\end{figure*}

\begin{deluxetable*}{cccccccc}
\tabletypesize{\small}
\tablewidth{0pc}
\tablecaption{\label{tbl:instrumentSpecs}\textit{Herschel} Observations}
\tablehead{
\ch {ObsID}	&\ch{Instrument}	&\ch{Band}		& \ch{Time on Target}& Scan Rate 		& \ch{Date(s)}	& \ch{Beam Size}		& \ch{Noise} \\ 
\ch{} 		& \ch{} 			&\ch{(\micron)}	& \ch{(s)}		& \ch{(\arcsec~s$^{-1}$)}	& \ch{Observed}	& \ch{(\arcsec)}		& (mJy) } 
\startdata
1342193149/50	& PACS  			& 100  			& 1129				& 20 				& 2010 Mar 31	&  6.69~$\times$~6.89		& 1.3 \\
1342193149/50	& PACS  			& 160  			& 1129				& 20 				& 2010 Mar 31 	& 10.65~$\times$~12.13		& 3.4 \\
1342220766/67	& PACS  			& 70   			& 1129				& 20 				& 2011 Apr 29 	&  5.46~$\times$~5.76		& 1.2 \\
1342220766/67	& PACS  			& 160			& 1129				& 20					& 2011 Apr 29 	& 10.65~$\times$~12.13		& 2.7 \\
1342204956	& SPIRE 			& 250 			& 901				& 30 				& 2010 Sep 21 	& 18.7~$\times$~17.5		& 4.9 \\
1342204956	& SPIRE 			& 350 			& 901				& 30 				& 2010 Sep 21 	& 25.6~$\times$~24.2		& 7.3 \\
1342204956	& SPIRE 			& 500 			& 901				& 30 				& 2010 Sep 21	& 38.2~$\times$~34.6		& 5.3
\enddata
\tablecomments{Beam sizes from PACS Observer's Manual v2.3 and SPIRE Observer's Manual v2.4.}
\end{deluxetable*}

Broadband photometric mapping observations were performed at 70, 100, 160, 250, 350, and 500\,$\mu$m using the Photoconductor Array Camera and Spectrometer (PACS; \citealt{PACS}) and Spectral and Photometric Imaging Receiver (SPIRE; \citealt{SPIRE}) instruments on board \Herschel~\citep{Herschel}. These observations were performed as part of the DEBRIS key programme (B. C. Matthews et al., in preparation) 
and were performed in mini and small scan-map modes for PACS and SPIRE, respectively. PACS observes at 100 and 160~\micron~simultaneously or at 70 and 160~\micron~simultaneously. Since \gamDor~was observed at both 70 and 100~\micron, \gamDor~was observed at 160~\micron~twice and the map at this wavelength is composed of the data from both observations. The coordinates of the peak of the emission in the 70 and 100~\micron~maps are used to align 160~\micron~maps. A summary of the observation parameters is given in Table~\ref{tbl:instrumentSpecs}. 

The PACS data were reduced using HIPE (\Herschel~Interactive Processing Environment: \citealt{2010Ott}) version 7.0 Build 1931. The reduction method includes some data that were taken while the telescope was reversing the scan direction, and therefore not scanning at a constant speed, to decrease the noise in the map. This reduces the noise at the map centre (where the target is located) by $\sim$25\%. PACS data are subject to 1/\textit{f} noise, whose power spectrum is a function of the noise at a given angular scale (determined by the scan rate of the telescope). To maximize the signal-to-noise in our maps, we filter out signals on angular scales larger than 66, 66, and 100\arcsec~at 70, 100, and 160~\micron~in the map generation process. The SPIRE data were also reduced using the standard \Herschel~pipeline script in HIPE.

Figure~\ref{fig:maps} shows the 60\arcsec\,$\times$\,60\arcsec~region around \gamDor~at 70, 100, and 160~\micron~and the 100\arcsec\,$\times$\,100\arcsec~region at 250, 350, and 500~\micron. A background source{,  which} is not listed in the NASA/IPAC Extragalactic Database (NED) or the NASA/IPAC Infrared Science Archive, is visible \bout 30\arcsec~to the southeast. The emission from \gamDor~and its disk is well resolved with PACS at 70, 100, 160~\micron~and marginally resolved at 250~\micron. The respective full width half maxima (FWHMs) of the star+disk observations at 70, 100, 160, 250 and 350~\micron~are 10.2\arcsec\,$\times$\,7.3\arcsec, 12.7\arcsec\,$\times$\,7.9\arcsec, 19.2\arcsec\,$\times$\,13.1\arcsec, 26.0\arcsec\,$\times$\,19.1\arcsec~and 27.5\arcsec\,$\times$\,22.8\arcsec~compared to the FWHMs for a standard star of 5.6\arcsec, 6.8\arcsec, 11.4\arcsec, 18.2\arcsec and 24.9\arcsec~(these are the geometric means of the beam sizes listed in Table~\ref{tbl:instrumentSpecs}). Emission is detected at all \Herschel~wavelengths, however, the background source to the southeast, which is well separated from the disk at 70--160~\micron, is harder to separate at SPIRE wavelengths. At 250 and 350~\micron, the nearby background source and \gamDor~begin to blend together, whereas at 500~\micron, the two are indistinguishable.

\section{\textit{Herschel} flux measurements}
\label{sec:fluxes}

Two methods are used to measure fluxes in the \Herschel~maps: aperture photometry is used when the emission is resolved and a point-spread function (PSF) is fit to the source when it is unresolved. Thus the flux from \gamDor~is measured with aperture photometry at 70--250~\micron, and PSF fitting at 350 and 500~\micron. The southeast background source is consistent with a point source in the maps, and so it is fit with a PSF at all wavelengths.
The PSF fit is done after rotating the instrumental PSF to match the rotation of the telescope at the time of the observations\footnote{The 160~\micron~PSF is composed of two equally weighted PSFs, each rotated to the corresponding position angle of the telescope at the time of the observation.} and using a $\chi^2$ minimization method. \Herschel~observations of the diskless star $\alpha$ Boo are reduced in the same manner as the DEBRIS data and used as the instrumental PSF at PACS wavelengths. Empirical SPIRE PSFs are downloaded from the ESA ftp site.\footnote{ftp://ftp.sciops.esa.int/pub/hsc-calibration/SPIRE/PHOT/Beams/}

\begin{deluxetable*}{ccccccl}[h!]
\tabletypesize{\footnotesize}
\tablewidth{0pc}
\tablecaption{\label{tbl:fluxes} Observed Fluxes and Predicted Photospheric Fluxes}
\tablehead{
\ch{Wavelength}	& \ch{Observed Flux}	&\ch{Instrument}& \ch{Method}	& \ch{Photosphere} & \ch{Excess} 	& \ch{Reference} \\
\ch{(\micron)}	& \ch{(Jy)}					& \ch{or Satellite} 			   			& \ch{} 	 		& \ch{(Jy)} 		  		& \ch{(mJy)} 	& \ch{} } 
\startdata
0.4		& 59.8 \p 0.8	& Hipparcos		&  \nodata 	 			& 59.7 \p 1.1	& \nodata 		& \cite{Hoeg2000} \\ 
0.5 		& 74.4 \p 0.8	& Hipparcos		&  \nodata 				& 74.8 \p 1.4	& \nodata		& \cite{Hoeg2000} \\ 
0.6 		& 71.5 \p 1.4	& Hipparcos		&  \nodata	 			& 71.7 \p 1.3	& \nodata		& \cite{1997ESASP1200P} \\ 
0.5 		& 76.5 \p 1.5	& Hipparcos		&  \nodata				& 75.7 \p 1.4	& \nodata		& \cite{2006yCat} \\ 
1.2 		& 53.7 \p 12.0	& 2MASS			&  \nodata				& 55.2 \p 1.0	& \nodata		& \cite{Cutri2003} \\ 
9.0 		& 1.908 \p 0.035	& AKARI			&  \nodata				& 1.924 \p 0.035	& \nodata		& \cite{Akari2010} \\ 
12.0 	& 1.202 \p 0.051	& \IRAS			&  \nodata 				& 1.179 \p 0.022	& \nodata		& \cite{1990IRASFC} \\ 
\hline
		& (mJy)			&				&						& (mJy)			&				&					\\
\hline
18.0 	& 514.2 \p 21.0		& AKARI		&  \nodata 				& 494.1 \p 9.1	& \nodata 		& \cite{Akari2010} \\ 
23.7 	& 315.6 \p 3.2		& MIPS		& PSF fit 				& 286.0 \p 5.3	& 29.6 \p 6.2 	& K. Su, private communication \\ 
25.0 	& 292.3 \p 22.0		& \IRAS		&  \nodata 				& 256.7 \p 4.7	& \nodata		& \cite{1990IRASFC} \\ 
60.0 	& 196.7 \p 11.0		& \IRAS		&  \nodata 				& 44.2 \p 0.8	& \nodata		& \cite{1990IRASFC} \\ 
71.4 	& 170.7 \p 8.1		& MIPS		& PSF fit 				& 31.1 \p 0.6	& 139.6 \p 8.1	& K. Su, private communication \\ 
70.0 	& 171.0 \p 8.7		& PACS		& 30 $\times$ 19\arcsec~aperture 	& 31.1 \p 0.6	& 139.5 \p 8.8	& This work \\
100.0 	& $<$476.7			& \IRAS		&  \nodata 				& 15.7 \p 0.3	& \nodata		& \cite{1990IRASFC} \\ 
100.0 	& 148.4 \p 7.7		& PACS		& 30 $\times$ 17\arcsec~aperture 	& 15.7 \p 0.3	& 132.7 \p 7.8	& This work \\ 
160.0 	& 134.3 \p 14.1		& PACS		& 30 $\times$ 20\arcsec~aperture 	& 6.4 \p 0.1		& 127.9 \p 14.1	& This work \\ 
250.0 	& 52.5 \p 6.5		& SPIRE		& 30 $\times$ 24\arcsec~aperture 	& 2.45 \p 0.05	& 50.0 \p 6.5	& This work \\ 
350.0 	& 23.5 \p 8.0		& SPIRE		& PSF fit 				& 1.24 \p 0.02	& 22.2 \p 8.0	& This work \\ 
500.0 	& $<$16.7 \p 5.9\tnm{a}	& SPIRE	& PSF fit 				& 0.60 \p 0.01	& $<$16.1 \p 5.9\tnm{a} & This work \\ 
\enddata 
\tnt{a}{Although there is a 3$\sigma$ detection at \gamDor's position at 500~\micron, the flux is listed as un upper limit since there is a known background source within the 500~\micron~beam.}
\end{deluxetable*}

At 250 and 350~\micron, where the background source is not well separated from the \gamDor~disk, two PSFs are fit simultaneously to the expected positions of each source. As a result, the disk flux at 350~\micron~is very uncertain (as well as the background source flux at 250 and 350~\micron). We quote the flux at 500~\micron~from a single PSF fit as an upper limit since the background source and the disk are within the same beam. The coordinates of the background source are determined from the 160~\micron~map, where it is both bright and well separated from the \gamDor~emission. The position of the PSF fit to \gamDor~is fixed to its expected coordinates at the time of the observations, given its proper motion (Table~\ref{tbl:gamdor}). This is reasonable as there is no evidence of an offset between the disk centre and the stellar position given that the location of the emission is 1.4 and 2.7\arcsec~from \gamDor's expected position at 70 and 100~\micron~(determined from the PSF fit) and therefore within \Herschel's 2.3\arcsec~pointing uncertainty (Herschel Observers' Manual v4).

The uncertainty in the flux is determined from the noise, listed in Table~\ref{tbl:instrumentSpecs}, measured by fitting a PSF to 400 random locations (the flux is the only free parameter) within a region devoid of significant emission and of similar coverage as the map centre (see B. C. Matthews et al., in preparation). (This is scaled by the area of the aperture for aperture-measured fluxes.) The PACS calibration accuracies of 3\%, 3\%, and 5\% at 70, 100, and 160~\micron, respectively (PACS Observer's Manual v2.3), and SPIRE pixel size correction factors and absolute flux calibration accuracy of 7\%~(SPIRE Observer's Manual v2.4) are added in quadrature. The measured fluxes of \gamDor~are listed in Table~\ref{tbl:fluxes}. The fluxes of the southeast background source are 11.4 \p 1.7, 20.0 \p 1.9, 35.0 \p 4.6, 28.0 \p 5.7, 18.5 \p 7.9, and $<$16.7 \p 5.9 mJy at 70, 100, 160, 250, 350, and 500~\micron. The 350~\micron~detection is only a 2.5$\sigma$ detection and the 500~\micron~3$\sigma$ detection (with respect to the map noise in Table~\ref{tbl:instrumentSpecs}) includes both \gamDor~and the background source.

Filtering out signals on large angular scales is necessary to reduce the noise (see Section~\ref{sec:obs} for more details), however, it also filters out the large angular scales of the PACS beam, which extends to about 17 arcmin. As a result, the fluxes measured in the PACS maps are too low. We correct for this by scaling our fluxes by 1.16 \p 0.05, 1.19 \p 0.05, and 1.21 \p 0.05 at 70, 100, and 160~\micron~\citep{Kennedy2012}. These correction factors are determined by comparing the fluxes of bright point sources in DEBRIS maps to their predicted photospheric flux. It should be noted that these correction factors are determined for point sources, but they are reasonable to use for an extended source, such as \gamDor, since the angular scales of the filtering are still large in comparison to the angular scale of \gamDor's emission.

\section{Photosphere and Excesses}
\label{sec:results}

\subsection{Modelling the Stellar Photosphere}
\label{sec:photmodel}

Accurate models of the stellar photosphere are crucial for debris disk studies, as the analysis is based on emission that is observed in excess of expected photospheric emission. The stellar photosphere is modeled using a $\chi^2$ minimization method to fit stellar models from the \textit{Gaia} grid \citep{Brott2005} to the observed optical and near-IR fluxes (see Table~\ref{tbl:fluxes}). The modeling of the stellar photosphere is done consistently for all DEBRIS targets (see \citealt{Kennedy2012} for more details). 

\begin{figure}
\includegraphics[trim=0.5cm 1cm 0cm 2cm, clip=true, width=3.5 in]{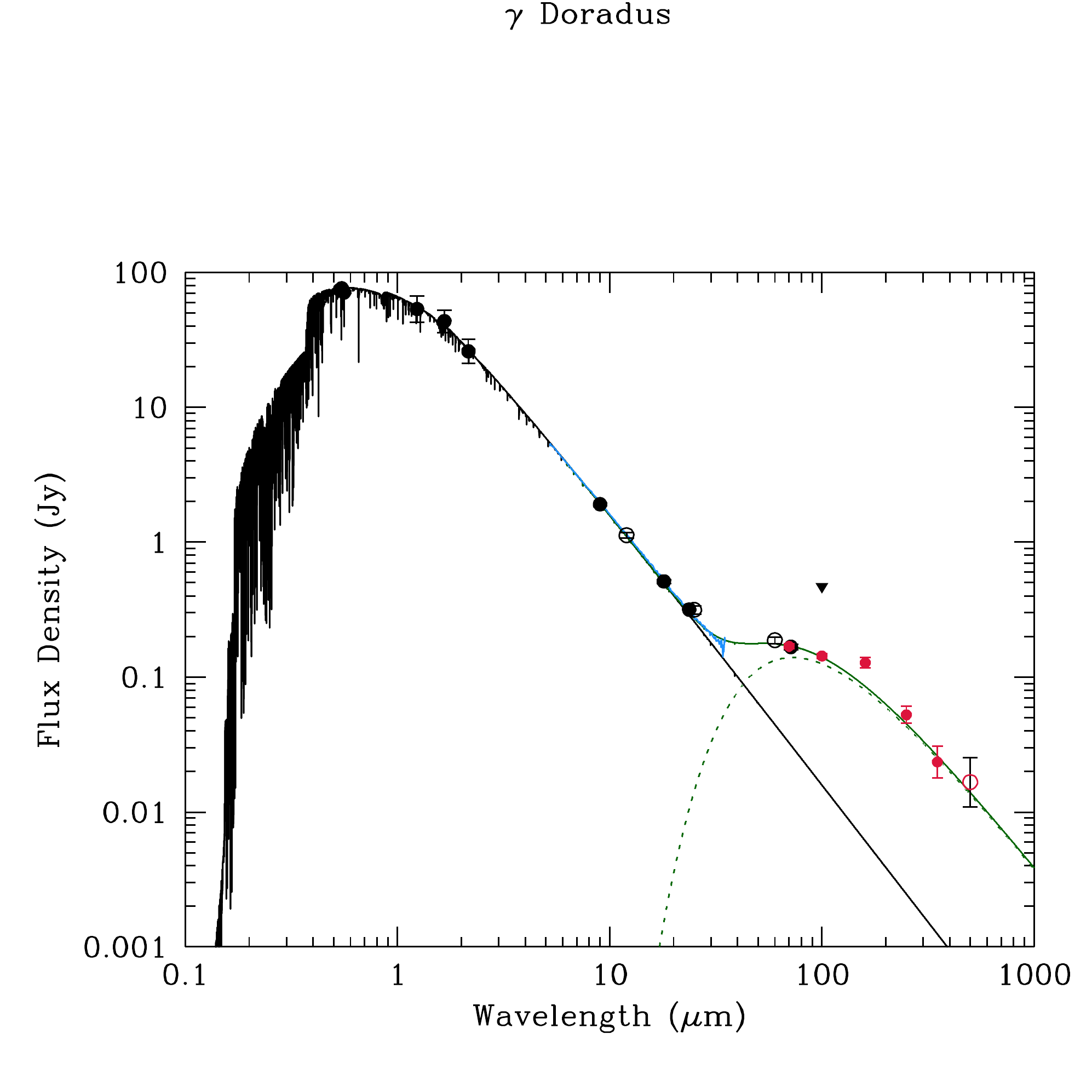}\centering
\caption{SED for \gamDor. \Herschel~fluxes are shown in \textit{red}. The IRS spectrum is plotted in \textit{blue}. Fluxes which contain a contribution from a background source (\IRAS~fluxes and our 500~\micron~flux) are displayed with \textit{open circles}. Upper limits are shown with \textit{triangles}. The \textit{black line} traces the photospheric model (Section~\ref{sec:photmodel}). The \textit{green dotted line} traces a sample 70~K blackbody spectrum to compare to the shape of the excess spectrum. The \textit{solid green line} traces the total flux from \gamDor~and the sample blackbody.}
\label{fig:SED}
\end{figure}

A photospheric model with a \Teff~of 7204 \p 28~K, a \logg~of 3.49 \p 0.1, an $[\rm{M/H}]$	 of $-0.41$ \p 0.17, an $R_*$ of 1.67 \p 0.22 R$_\odot$ and an $L_*$ of 6.7 \p 0.1 L$_\odot$ is used. Given the uncertainties in deriving \logg~and [M/H] from SED modeling, the results are in good agreement (within 0.1 dex and 0.3 dex, respectively) of \cite{GrayCorbally2006}.  The predicted photospheric fluxes from the stellar fit are compared to the observed fluxes and used to measure the excess emission in Table~\ref{tbl:fluxes}. The SED is shown in Figure~\ref{fig:SED} with the photospheric model.

\subsection{\Spitzer~Excesses}
\label{sec:irs}

We include \Spitzer~Infrared Spectrograph (IRS) observations of \gamDor~downloaded from the Cornell Atlas of \Spitzer/IRS Sources (CASSIS). This spectrum is from \cite{Ardila2010} and includes AOR 24368640, AOR 27577600, and AOR 3555584. 
It is consistent with no excess shortward of 22~\micron~and is therefore in agreement with AKARI fluxes that measure no excesses shortward of 18~\micron.

\section{{Basic image analysis}}
\label{sec:basicimanal}

\begin{deluxetable*}{cccccccc}
\tabletypesize{\footnotesize}
\tablewidth{0pc}
\tablecaption{\label{tbl:gaussianFits} 2D Gaussian Disk Fits}
\tablehead{
\ch{Band}	& \ch{Beam\tnm{a}} & \ch{$\theta_{\rm{obs,maj}}$} & \ch{$\theta_{\rm{obs,min}}$} & \ch{$\theta_{\rm{act,maj}}$} & \ch{$\theta_{\rm{act,min}}$}& \ch{Inclination} & \ch{Position Angle} \\ 
\ch{(\micron)} &\ch{(\arcsec)} & \ch{(\arcsec)} & \ch{(\arcsec)} & \ch{(\arcsec)} & \ch{(\arcsec)} & \ch{(\degree)} & \ch{(\degree)} }  
\startdata
70		& 5.61		& 12.7 \p 0.7	& 7.9 \p 0.4			& 11.4 \p 0.8		& 5.6 \p 0.6 	& 61 \p 6 	& 56 \p 6 \\ 
100		& 6.79		& 14.2 \p 0.7	& 8.2 \p 0.4			& 12.5 \p 0.8		& 4.6 \p 0.7 	& 68 \p 5 	& 52 \p 4 \\  
160		& 11.36		& 20.2 \p 0.8	& 13.3 \p 0.5		& 16.6 \p 1.0		& 6.9 \p 1.0 	& 65 \p 5 	& 62 \p 5 \\ 
250		& 18.2		& 26.3 \p 4.2	& 19.3 \p 3.1		& 19.0 \p 5.8		& \nodata \tnm{b} 	& $>$52\tnm{b}	& 55 \p 22 \\ 
350		& 24.9		& 26.3 \p 6.9	& 22.7 \p 6.8		& \nodata \tnm{c}			& \nodata \tnm{c}		& \nodata \tnm{c}		& \nodata \tnm{c} \\
\enddata
\tnt{a}{The effective beam size (geometric mean) is listed and used to deconvolve the sizes using Equation~(\ref{eqn:convolve}).}
\tnt{b}{The disk is only marginally resolved along one axis at 250~\micron. Therefore $\theta_{\rm{act,min}}$ cannot be derived for the short axis and consequently only the upper limit on the inclination can be calculated.}
\tnt{c}{The system is consistent with being unresolved and symmetric at 350~\micron.}
\end{deluxetable*}

The size of the disk can be estimated by fitting 2D Gaussians to the images. First, a model of the photosphere is removed from the image by subtracting a PSF  that is scaled to the expected photospheric emission (Section~\ref{sec:photmodel}). The FWHMs and position angles of the 2D Gaussian models are listed in Table~\ref{tbl:gaussianFits}. The quoted error on a fitted parameter gives the range of parameter values for which the \chisq~value is within 10\% of the minimum \chisq~value. The \chisq~value is  measured in a small region around \gamDor.\footnote{30{\arcsec}\,$\times$\,30{\arcsec}, 30{\arcsec}\,$\times$\,30{\arcsec}, 40{\arcsec}\,$\times$\,40{\arcsec}, 70{\arcsec}\,$\times$\,70{\arcsec}, and 80{\arcsec}\,$\times$\,80\arcsec regions are used at 70, 100, 160, 250, and 350~\micron, respectively.} 

The major and minor FWHMs are deconvolved from the beam size using 
\begin{equation}
\label{eqn:convolve}
\theta_{\rm{obs}}^2 = \theta_{\rm{act}}^2 + \theta_{\rm{beam}}^2,
\end{equation}
where $\theta_{\rm{act}}$ is the actual angular size of the object, $ \theta_{\rm{beam}}$ is the effective beam size and $\theta_{\rm{obs}} $ is the observed  angular size of the object. (This is the relation for convolving a Gaussian with a Gaussian.) Assuming the true shape of the disk is azimuthally symmetric, the deconvolved major and minor FWHMs are used to estimate an inclination of $\sim$65\degree~from a face-on orientation (see Table~\ref{tbl:gaussianFits}) that is consistent with the more detailed image modeling (see Section~\ref{sec:imagemodel}). 

\begin{figure}
\includegraphics[trim=1cm 0cm 1.75cm 0cm, clip=true, width=3.5 in]{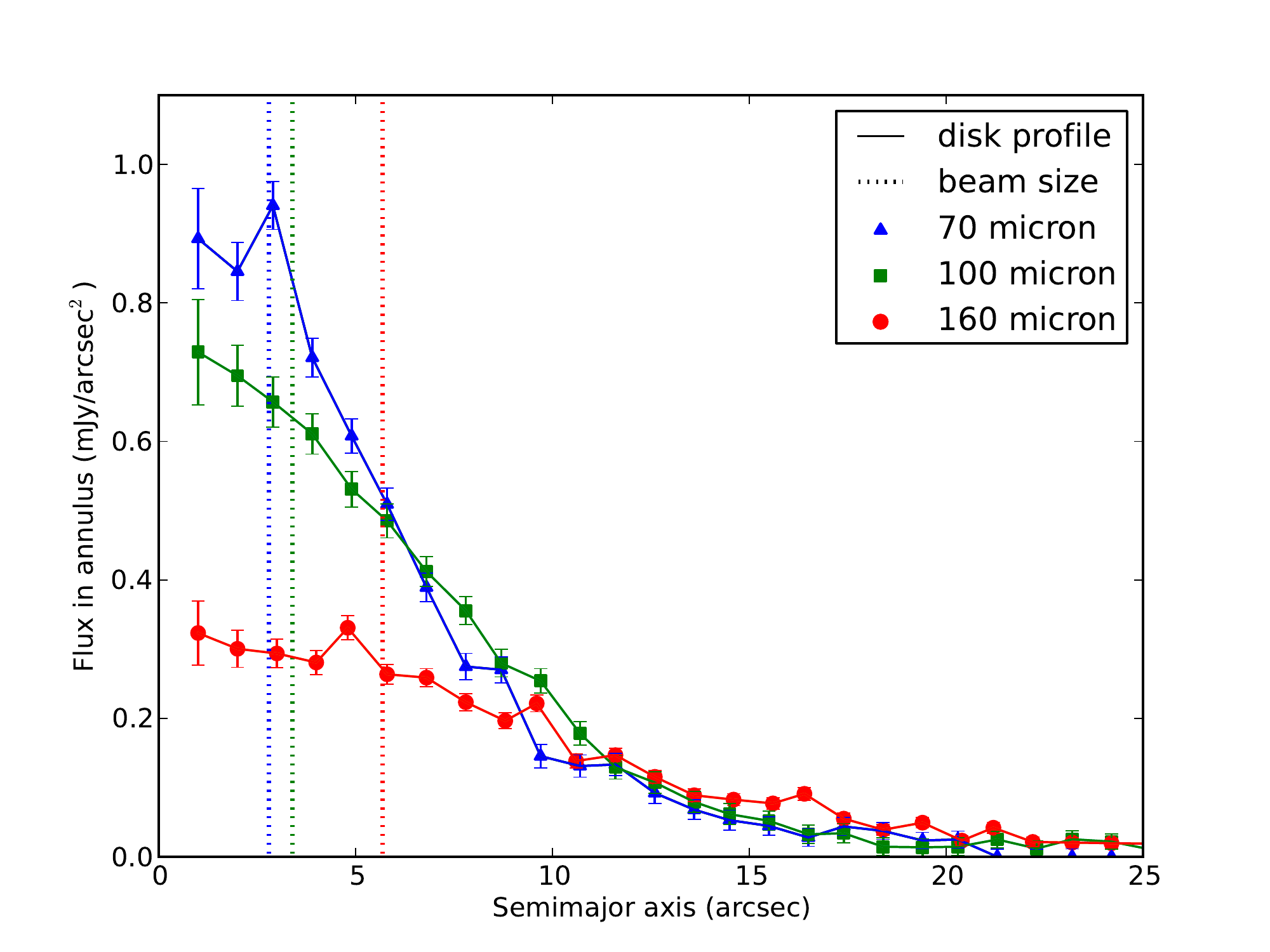}\centering
\caption{Surface brightness profiles of the \gamDor~disk. The average surface brightness in elliptical annuli with increasing semi-major axis at 70 (\textit{blue triangles}), 100 (\textit{green squares}) and 160~(\textit{red circles})~\micron. 1$\sigma$ errors are shown. The stellar model is subtracted from the maps to measure the profile of the disk (\textit{solid lines}). (The model of the nearby background source is also subtracted from the maps.) Beam sizes (\textit{dotted lines}) are plotted for comparison. The shape of the profile is consistent with a radial distribution of dust where the cooler dust is brighter and the warmer dust is fainter at longer wavelengths (see the text). The surface brightness profiles fall off to similar levels beyond \bout10\arcsec~at all wavelengths. }
\label{fig:fluxDistribution}
\end{figure}

The fitted parameters of the 2D Gaussian models are used to define elliptical annuli to measure the surface brightness profiles of the disk in the 70, 100, and 160~\micron~images where the disk is resolved along both the major and minor axes. Figure~\ref{fig:fluxDistribution} shows that the shape of the profiles is wavelength dependent. This is consistent with a broad radial distribution of material as dust further from the star will be cooler and dominate the flux at longer wavelengths. This is also reflected in the larger FWHMs at longer wavelengths listed in Table~\ref{tbl:gaussianFits}. The underlying dust distribution is investigated by modeling the images in the following section.

\section{Disk modeling}
\label{sec:modeling}

\subsection{Basic Model}
\label{sec:basicmodel}

We first present the underlying set up of the debris disk modeling that we implement in Sections~\ref{sec:SEDmodel} and \ref{sec:imagemodel}, to introduce the various parameters. The simple approach to modeling debris disks is to assume that the dust is contained within a narrow/discrete ring at some distance from the host star. This model is easily extended to model multiple narrow/discrete rings by summing up the individual contributions of each ring using a modified blackbody function \citep{Dent2000},
\begin{equation}
\label{eq:modbb}
F_\nu = 2.35 \times 10^{-11}~d^{-2}~\sum_R~ \sigma(R)~B_\nu (\lambda, T(R))~X_{\lambda}^{-1}
\end{equation}
where $F_\nu$ is the flux density (in Jy) at wavelength $\lambda$ (in~\micron), $d$ is the distance to the star (in pc), and $\sigma(R)$ is the cross-sectional area (in~AU$^2$) of the ring at radius $R$ (in~AU). $B_\nu(\lambda, T(R))$ is the Planck function (in Jy sr$^{-1}$) for the ring at radius \textit{R} with temperature \textit{T} (in~K). Equation~(\ref{eq:modbb}) is modified by $X_\lambda$, where $X_\lambda = (\lambda/\lambda_0)^\beta$ for wavelengths longer than  $\lambda_0$ and  $X_\lambda =1$ otherwise. This accounts for the fall off of the spectrum that is observed to be steeper than the blackbody function at submm wavelengths for most debris disks \citep{Dent2000}.

If the dust grains act like blackbodies, the radius of a dust ring can be derived from its temperature given the luminosity of the star, $L_*$:
\begin{equation}
\label{eq:Rbb}
{R_{bb}} = 278.3^2 ~L_*^{1/2}~T^{-2}.
\end{equation}
However, using the above method to model the SEDs of debris disks typically underestimates the observed sizes of the disks by a factor of 2--5 (\citealt{Schneider2006,Wyatt2008,Matthews2010,RodriguezZuckerman2012}; Booth et al., accepted, and references therein) because grains with a lower radiation efficiency, dependent on their size and composition, will be hotter than blackbody grains at the same distance would.

Equation~(\ref{eq:modbb}) can be expanded to model dust in a wide/extended belt by treating the belt as a series of dust rings and setting $\sigma$ and $T$ to follow power-law distributions. The cross-sectional area is determined from the optical depth, $\tau(R)$, given by
\begin{equation}
\label{eq:powerlawsigma}
\tau(R) = \tau_0~(R/R_0)^{-\gamma},
\end{equation}
where $\tau_0$ is the optical depth at $R_0 = 1$~AU and $\gamma$ describes the fall off of the surface density at further distances from the star. 
Similarly, $T(R)$ is given by,
\begin{equation}
\label{eq:powerlawtemp}
T(R) = T_0~(R/R_0)^{-\delta},
\end{equation}
where $T_0$ is the temperature at $R_0 = 1$~AU and $\delta$ describes how quickly the temperature of the dust grains declines at larger radii from the star. The blackbody temperature distribution given in Equation~(\ref{eq:Rbb}) has the values of $\delta = 0.5$ and $T_0 = 438.4$~K for \gamDor. The cross-sectional area, $\sigma$, the fractional luminosity, $f_d$, and the optical depth, $\tau$, of the dust are related by $\sigma(R)~=~2\pi\,R\,{dR}\,\tau(R)\,=\,f_d\,4 \pi\,R^2$ (see \citealt{Krivov2010} for a review). For a narrow/discrete ring model, we assume $dR/R = 0.1$.

This construct is valid for modeling both the SED and the resolved images. The SED modeling traces the temperatures of the dust within the disk (e.g., $T$, $\sigma$), whereas the image modeling constrains its physical arrangement ($R$ and therefore the dependence of $T$ and $\sigma$ on $R$).
This approach (used for all DEBRIS modeling papers) parameterizes the shape of the SED at each radius. How that shape relates to the physical parameters of dust composition and size distribution will be discussed in a later paper. 

\subsection{SED Modelling}
\label{sec:SEDmodel}

For the majority of debris disks, only unresolved photometry is available. We include an analysis of the SED without the resolved spatial information to highlight how the resolved images reduce the ambiguities. A single dust temperature is unable to account for the spectral breadth of the excess emission (Figure~\ref{fig:SED}). We therefore model the SED with two different extended temperature distributions: a model of two discrete dust rings and a model of a single extended dust belt. All models have the same fractional luminosity of $2.6 \times 10^{-5}$.

The SED model of two discrete rings accounts for the broad shape of the SED with a warm ring ($T$ \bout~125~K, $R_{bb}$ \bout~12~AU) that dominates the flux at IRS wavelengths and a cool ring ($T$ \bout~50~K, $R_{bb}$ \bout~77~AU) that accounts for the \Herschel~fluxes. This model implies that the \Herschel~images would be dominated by dust at a single temperature.

The extended belt SED model requires a relatively flat profile of the cross-sectional area ($\gamma \sim 0$ in Equation~(\ref{eq:powerlawsigma})) to account for the similar fluxes at 70, 100, and 160~\micron~and the mid-IR (IRS and MIPS 24) excesses. Using a blackbody temperature distribution, this model suggests that the dust extends from \bout13~AU to \bout166~AU. 

\begin{figure*}
\includegraphics[width=6.5in, clip=True, trim= 0.4cm 12.9cm 0.3cm 7.9cm ]{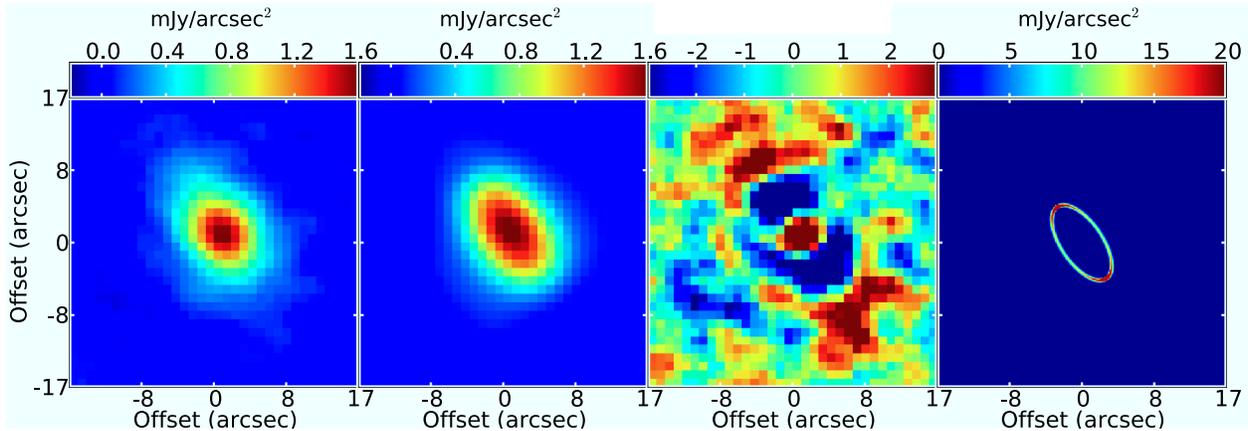}\centering
\caption{An attempted single narrow ring model of the \gamDor~disk. \textit{Left to right}: the observations, the simulated \Herschel~observations of the model, the residuals, and simulated observations of the model at high resolution at 70 \micron. The color scale of the residuals extends from $-3\sigma$ to 3$\sigma$. Clearly the single narrow ring model cannot adequately explain the observed emission and we must turn to a model that includes dust at a range of radii.}
\label{fig:img_1ring}
\end{figure*}

The immediate contrast between unresolved and resolved photometry is evident in the predictions from the SED models. The SED model of two discrete rings suggests that the \Herschel~fluxes are dominated by a single narrow ring at 77~AU, however, the resolved images show an extended distribution of dust that cannot be modeled with a single ring, as shown in Figure~\ref{fig:img_1ring}. 
The discrepancies between what would be interpreted from the SED and what is revealed by modeling resolved images (see the following section) are clear. Resolved imaging is the only way to constrain the possible spatial distributions.

\begin{deluxetable}{cc}
\tablewidth{0pc}
\tablewidth{3in}
\tablecaption{\label{tbl:img_tworing} Fitted Parameters for the Two Narrow Rings\\ \textcolor{white}{wwhhiitte} Model of the \Herschel~Images}
\tablehead{
\ch{Parameter}		& \ch{Value}} 
\startdata
$R_1$				& 70~AU \\
$R_2$ 				& 190~AU \\
$\tau_{2-1}$			& 0.85 \\
Inclination 			& 71\degree \\
Position angle 		& 55.9\degree~N of E \\
70~\micron~flux 		& 140~mJy \\
100~\micron~flux 	& 142~mJy \\
160~\micron~flux 	& 125~mJy \\
\enddata
\end{deluxetable}

\subsection{Image Modelling}
\label{sec:imagemodel}

To keep the results from SED modeling and image modeling distinct, we describe our imaging models as ``narrow rings" and ``wide belt" which are analogous to the ``discrete rings" and ``extended belt" SED models, but of course with different fit parameters. We also refer to the dust observable to \Herschel~as ``cool dust'' and dust that is responsible for any excesses at shorter wavelengths (IRS and MIPS 24) that does not effect the \Herschel~observations as ``warm dust.''

The extended spatial distribution of dust is clear in the images. The best-fit imaging model of a single narrow dust ring is unable to account for the emission on the largest and smallest scales (see Figure~\ref{fig:img_1ring}). Although the images show that all the dust cannot lie at the same stellocentric radius, the resolution of \Herschel~is not sufficient to distinguish between different extended configurations. Below, we show that both a model of two narrow rings of dust and a model of a wide belt of dust are able to reproduce the observations of the cool dust component. Both models are parameterized by an inner radius and an outer radius. (A constant opening angle of 10\degree~was used and reasonable variations to this do not affect the fitting.) The narrow rings model describes two discrete dust components at these radii with no dust between them whereas the wide belt model is described by a smooth distribution of material between these edges. We do not intend to fully investigate the possible parameter space given the limited resolution. Rather, we compare the two models to demonstrate the uncertainties on the radial extent of the disk. These two extended spatial distributions are reasonable to consider as they are observed in other resolved debris disks (e.g., $\eta$ Corvi: \citealt{Wyatt2005etaCorvi,Matthews2010}, HR 8799: \citealt{Su2009}, HD 181327: \citealt{Schneider2006}). The two configurations are modeled with similar techniques at 70, 100, and 160~\micron, where the \gamDor~disk is well separated from the nearby background source. A \chisq~grid analysis is used for the two narrow rings model (as in Booth et al., accepted), and a combination of by-hand and \chisq~minimization (as in \citealt{Kennedy2012} and \citealt{Wyatt2012}) is used for the wide belt model. 

\begin{figure*}
\includegraphics[trim=0.67cm 8.5cm 0.5cm 3.5cm, clip=true, width=7 in]{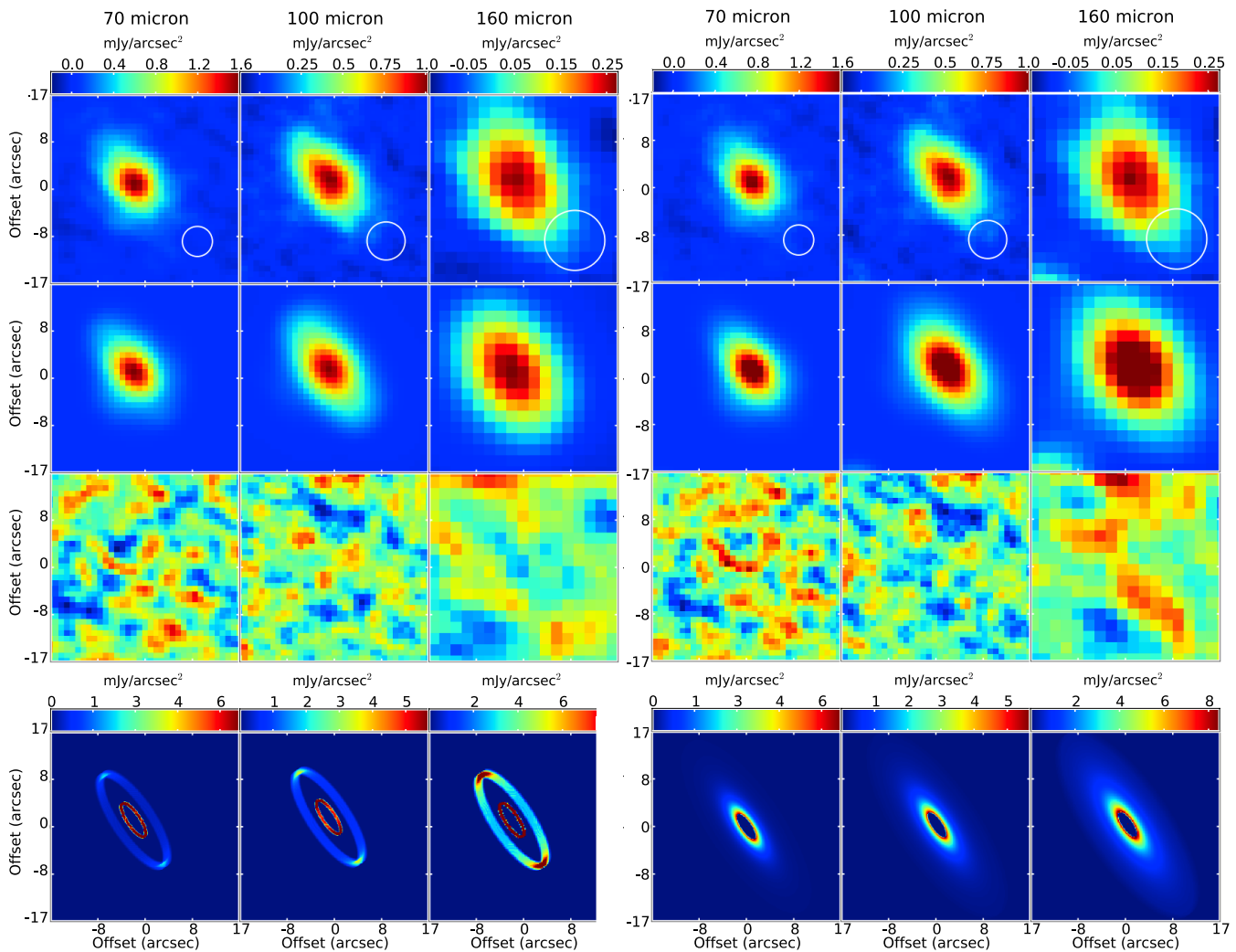}\centering
\caption{Imaging models of the \gamDor~disk. (\textit{top to bottom}) The observations, the simulated \Herschel~observations of the model, the residuals, and the simulated observations of the model at high resolution at (\textit{left to right}) 70, 100 and 160~\micron~for the imaging models of the cool dust. The narrow rings model is on the left, and the wide belt model is on the right (see the text). The color scale for the residuals map extends from $-3\sigma$ to $+3\sigma$. The residuals show that both models of the cool dust successfully reproduce the resolved PACS observations.}
\label{fig:img_models}
\end{figure*}

\begin{deluxetable}{cc}
\tablewidth{3in}
\tablecaption{\label{tbl:img_wide} Fitted Parameters for the Wide Belt Model \\ \textcolor{white}{wwhhiitte}  of the \Herschel~Images}
\tablehead{
\ch{Parameter}		& \ch{Value} } 
\startdata
Inner radius			& 55~AU \\
Outer radius			& 400~AU \\
$\tau_0$				& $3.4 \times 10^{-5}$ \\
$\gamma$				& 0.8 \\
$T_0$				& 585~K \\
$\delta$\tnm{a}		& 0.5 \\
Inclination			& 68.5\degree \\
Position angle		& 55.6\degree~N of E \\
\enddata
\tnt{a}{These parameters are fixed in the model.}
\end{deluxetable}

The best-fit imaging model of two narrow rings is determined using a grid of parameters: the radius of the inner ring ($R_1$), the radius of the outer ring ($R_2$), the inclination of the disk, and the ratio of the optical depths of the two rings ($\tau_{2-1} = \tau_1/\tau_2$, where $\tau_1$ and $\tau_2$ are the $\tau$ for the inner and outer rings, respectively).\footnote{The grid contains 225,000 models with radii from 55 to 330~AU and $\tau_{2-1}$ from 0.1 to 1.5.} The grid size (and therefore computation time) is minimized by fixing the position angle of the disk to that determined from the 2D Gaussian model of the 70~\micron~image (Section~\ref{sec:results}). Additionally, for each set of grid parameters, the total disk flux, and therefore the total $\tau$ of the system, is fit using the package MPFIT \citep{Markwardt2009}. The model and its residuals in Figure~\ref{fig:img_models} show that the two narrow rings model is able to reproduce the \Herschel~observations. The fitted parameters for this model are listed in Table~\ref{tbl:img_tworing}.

The wide belt imaging model is parameterized by an inner radius, a disk width, a power-law surface density with slope described by $\gamma$ (Equation~(\ref{eq:powerlawsigma})), an inclination to the line of sight and a position angle on the sky. The temperature of the dust grains is modeled to fall off with $\delta = 0.5$ (Equation~(\ref{eq:powerlawtemp})) as it was not necessary to deviate from this power law. The images are modeled using a Levenberg--Marquardt \chisq~minimization. This technique runs the risk of finding a local minimum in the \chisq~value, however, it is less computationally intensive than using a grid and the best-fit model reproduces the images, and so must be considered a plausible representation of the disk structure. The model and its residuals are shown in  Figure~\ref{fig:img_models} and the  fitted parameters are listed in Table~\ref{tbl:img_wide}. The fitted $T_0$ (585~K) is hotter than that for a blackbody (438~K). This is likely due to the dust grain composition and/or size distribution. 

We compute the reduced \chisq~of each model in the 49\arcsec\,$\times$\,49\arcsec, 49\arcsec\,$\times$\,49\arcsec, and 50\arcsec\,$\times$\,50\arcsec~region\footnote{The region around the background source is not included in this calculation as each model treats the background source differently.} around \gamDor~at 70, 100, and 160~\micron. The residuals of both models are reasonably low with $\chi^2_{dof} =$ 0.78 and 0.71 for the narrow rings and wide belt models, respectively. This suggests that both configurations are possible for the cool dust around \gamDor. We use the best fits of each model to estimate the uncertainty on the radial extent of the disk. The narrow rings model shows the significant emission extends out to at least 190~AU. Emission beyond that is predicted by the wide belt model to extend out  to 400~AU, however it is difficult to constrain given its low surface brightness. This makes the outer radius of the wide belt model highly uncertain. There is also a discrepancy of approximately 15~AU between the inner extent of each model. This demonstrates that the dust visible to \Herschel~has an inner radius around 60~AU, but cannot be constrained to better than 25\%. The resolution of this disk is sufficient to place much tighter constraints on the geometrical viewing parameters of the disk. The inclinations of the imaging models (68.5\degree~and 71\degree) are in good agreement and both models measure a position angle of 56\degree.

\begin{figure}
\includegraphics[trim=0.75cm 0cm 1.75cm 1.25cm, clip=true, width=3.5 in]{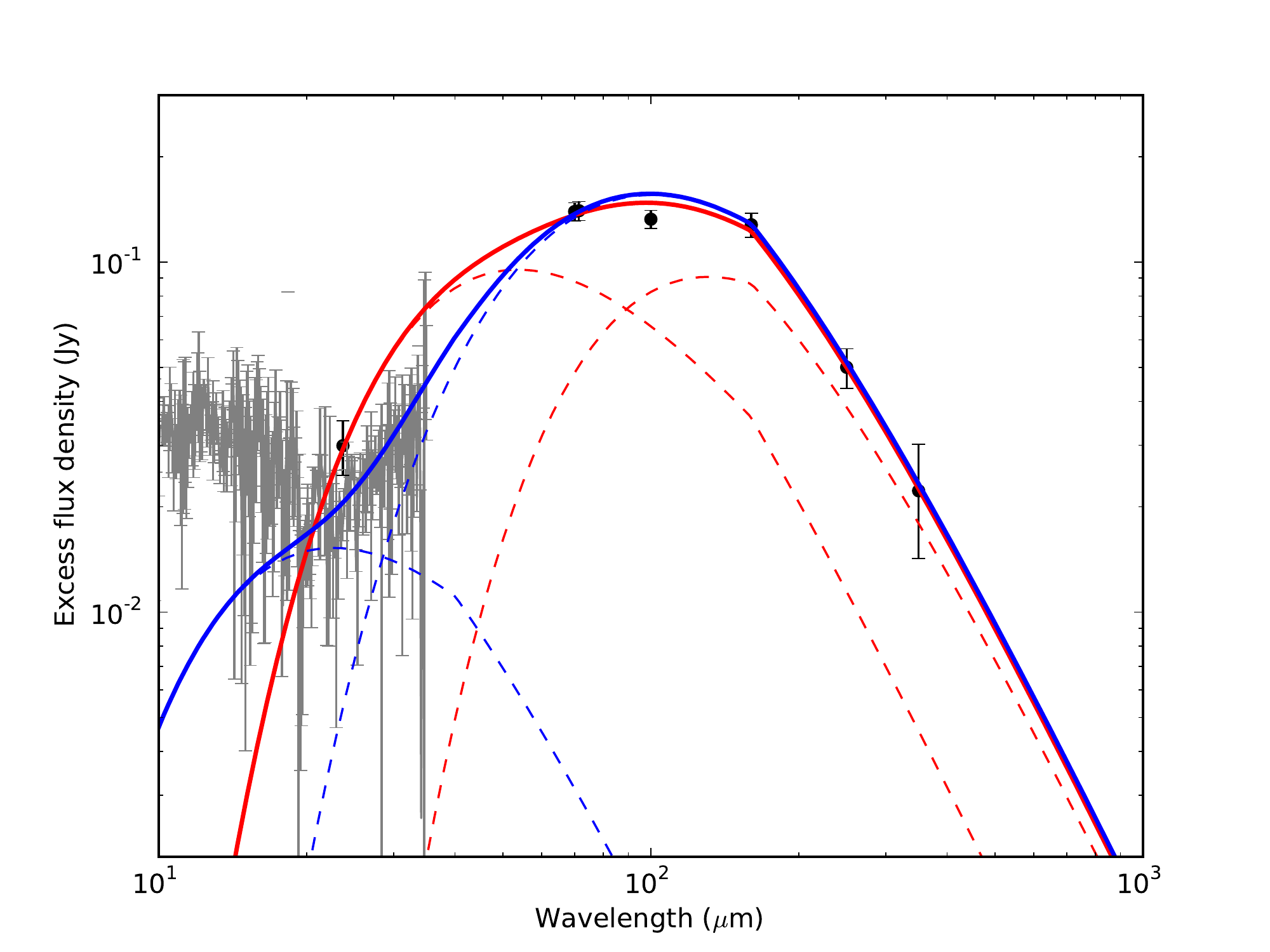}\centering
\caption{SED of the \gamDor~disk derived from the imaging models. Photometric excesses are displayed with black circles and IRS excesses are in grey. The narrow rings imaging model is plotted in red and the wide belt imaging model is plotted in blue. Individual components of each model are plotted with dashed lines and the solid lines trace the total flux of each disk model. Both models require two components to reproduce the observations. The narrow rings model agrees well with the MIPS 24~\micron~data and places the rings at \bout70 and \bout190~AU. The wide belt imaging model that extends from \bout55~AU to \bout400~AU cannot account for the excesses at shorter wavelengths. Therefore, an additional component, a warm, narrow ring with a temperature of \bout 225 \p 100~K,  is needed and its properties are estimated from the unresolved mid-IR fluxes.}
\label{fig:components}
\end{figure}

Figure~\ref{fig:components} shows the SEDs generated for the imaging models. We have adopted the standard $\beta = 1$ that is found for most debris disks \citep{Dent2000} and fixed $\lambda_0 = 160$~\micron~to accommodate the SPIRE observations. Both the narrow rings and the wide belt imaging models overpredict the observed 100~\micron~flux. This discrepancy is also present in the SED models (Section~\ref{sec:SEDmodel}) and therefore not an issue with the imaging models themselves. The apparent flatness of the SED between 70 and 160~\micron~is not characteristic of blackbody emission. More sophisticated models could explore whether this feature is due to the dust's composition, causing broad emission features from minerals or ices, or due to the shape of the size distribution. For example, \cite{Lebreton2012} show different SED shapes for HD 181327's debris disk for various grain compositions and size distributions.

The SED of the wide belt imaging model suggests that the excess at shorter IRS and MIPS 24 wavelengths is due to a separate component of warm dust that does not contribute to the emission at \Herschel~wavelengths, rather than an extension of the belt. We therefore attribute the remaining warm excess in the context of this model to a separate warm component of dust and estimate it to have a temperature of 225~K and a fractional luminosity of $4.4 \times 10^{-6}$.  In order to minimize the warm component's contribution to the flux at \Herschel~wavelengths, we adopt a $\lambda_0$ of 40~\micron. The parameters of this component are highly uncertain given the low significance of excess and the fluctuation of the degree of excess between different IRS reductions, and so we estimate an uncertainty of \bout 100~K on the dust temperature. Given that the excess is apparent in all available reductions of the IRS data and MIPS photometry at 24~\micron, which is even more discrepant, we are confident that it exists. The 225~K dust temperature suggests that the warm inner ring lies at \bout4~AU under blackbody assumptions using Equation~(\ref{eq:Rbb}). The SED modeling supports that such a warm dust temperature is present even though its physical location is poorly constrained and could lie anywhere between 2 and 12~AU given the uncertainty in temperature. Typically, the blackbody radius underestimates the true size of the disk by a factor of up to \bout5 (\citealt{Schneider2006,Wyatt2008,Matthews2010, RodriguezZuckerman2012}; Booth et al., accepted) 
implying the true radius could exceed the nominal blackbody radius even further.

Modelling the SED and resolved images suggests that there are at least two dust components in \gamDor's debris disk. However, the resolution of the images and the SEDs are unable to distinguish between a configuration of two narrow dust rings (at 70 and 190~AU) and a configuration of a wide outer dust belt (from 54 to 400~AU) and a warm inner ring (around 4~AU).

\section{Discussion}
\label{sec:discussion}

\subsection{Properties of \gamDor's Disk}

It is important to constrain where the dust is to determine the implications for a possible planetary system (Section~\ref{sec:planetary_systems}). The modeling of the resolved images shows that the dust population probed at \Herschel~wavelengths could be distributed in either a wide belt and a narrow ring or two narrow rings. Figure~\ref{fig:components} shows that both models have two dust components. The narrow rings model requires a gap between the star and the inner narrow ring at \bout70~AU and another gap between the inner ring and the outer narrow ring (at \bout190~AU). Whereas the wide belt model requires a gap between a narrow ring at \bout4 and the wide belt that begins at \bout55~AU. All models have an estimated fractional luminosity of \bout$10^{-5}$, within the upper limits that are typically found for other debris disks around F-type stars \citep{Moor2011}.

A common clearing mechanism that is proposed for such gaps, for example, is the presence of planets, however, they would be hard to form at such large distances from the star (e.g., \citealt{Weidenschilling2003}). Therefore, the wide belt is physically more plausible since it does not require planets at such large radii. 
However, there are problems with the wide belt model as well, as it would require a very broad disk of dust producing planetesimals, rather than containing them in narrow rings.

The inclination of the disk is \bout69\degree~from a face-on orientation. This suggests that it may be aligned with the stellar equator given the inclination of \gamDor~itself is \bout70\degree~\citep{Balona1996}, though the orientation of the stellar inclination is not known.

The fractional luminosity of the warm inner ring of \bout $4.4 \times 10^{-6}$ is about the maximum that can be expected to arise from collisional processes for Sun-like stars from \citet{Wyatt2007} 
\begin{equation}
\label{eq:fmax}
f_{\rm{max}} = 0.16 \times 10^{-3}~R^{7/3}~M_{*}^{-5/6}~L_{*}^{-1/2}~t_{\rm{age}}^{-1},
\end{equation}
where $R$ is the dust radius in~AU, $t_{\rm{age}}$ is the age of the star in Myr, $M_{*}$ is the mass of the star in solar masses, and $L_{*}$ is the luminosity of the star in solar luminosities (Section~\ref{sec:photmodel}).
Given \gamDor's age of 400 Myr, $f_{\rm{max}}$ is $2.5 \times 10^{-6}$ for a dust ring at 4~AU and therefore transient events are not needed to explain this amount of hot dust. 


\subsection{{Relation to Planetary Systems}}
\label{sec:planetary_systems}

\gamDor~is an example of how we are continuing to uncover more structure in debris disks with the resolution and sensitivity of modern observatories to move away from the simple pictures of single narrow dust rings. Observing \gamDor~with \Herschel~demonstrates the importance of surveys across different wavelength regimes. In the case of \gamDor, the \Spitzer~and \Herschel~observations considered independently would give an incomplete picture of the system. Our understanding of the debris disk around \gamDor~now is that it is more complex than a single ring. We find that it is composed of dust at multiple radii 
by showing two extreme scenarios of this are both possible: two cool, narrow rings beyond 70~AU or a cool, wide belt beyond 55~AU and a separate warm inner ring.
In both cases the dust is located between \bout55~AU and a few hundred AU. Depending on the exact scenario (particularly in the case of a wide belt), an additional warm component of dust (at several AU from the star) may be required to reproduce the mid-IR excess. 
The wide belt model with a warm inner component is much like the Solar System, although it is much younger than the Solar System, the dust is much brighter and no planets in the system have been detected yet. 

\gamDor's multicomponent disk has implications for the possibility of the system hosting planets. For planetary systems, such as the Solar System and HR 8799, 
that host both planets and a debris disk, the planets are observed to lie within or between the regions of dust \citep{MoroMartin2010}. This suggests that gaps in debris disks are good places to look for planets, particularly if the dust ring has a sharp inner edge \citep[e.g., Fomalhaut; ][]{Kalas2005,Kalas2008}. Resonance overlap studies have shown how a planet can sculpt the inner edge of a disk (e.g., \citealt{Quillen2006}, \citealt{MustillWyatt2012}).
Migrating planets as well as planets with highly eccentric orbits can create gaps and asymmetries in debris disks (see e.g., \citealt{Wyatt2003}). 
\gamDor~is a good target for the Atacama Large Millimeter/Submillimeter Array (ALMA) as \gamDor's distance is appropriate to resolve a sharp inner edge. Such observations would constrain the location of edges and therefore help distinguish between the two models presented here. If \gamDor's disk is indeed best described by a wide belt, the best place to look for planets would be within $\sim$55~AU of the star, where the gap between the inner ring and the wide belt is predicted. Similarly in the context of the narrow rings model a planet could be sculpting the inner ring at 70~AU and/or planets could be responsible for the gap between 70 and 190~AU.

Determining the inner edge of the disk could also have implications for the histories of the planetary systems. For example, \cite{Wyatt2003} present a model of a Neptune-mass planet migrating from 40 to 65~AU. 
During the migration, the inner edge of the disk is pushed outward and objects in the disk are swept into resonances with the planet. Migration of the Solar System planets has been proposed as a possible trigger for the Late Heavy Bombardment (LHB), an epoch of intense cratering responsible for many of the features we still see on the Moon today. The LHB severely depleted the Solar system's debris disk rendering such a system unobservable at the distance of \gamDor~within the current detection limits and available observatories \citep{Booth2009}. Although it would be unlikely to observe a planetary system while its planets are migrating, because of the short timescales over which this takes place (on the order of tens of Myr), the inner edges of debris disks and the amount of dust within them are tools to consider the Solar System's history in the context of other planetary systems. For example, the fact that there is so much dust still around \gamDor~suggests that it has not undergone an instability as destructive as the LHB.

\section{Summary and conclusions}
\label{sec:conclusions}

We observed the debris disk around \gamDor~with \Herschel~and detect emission at all six wavelengths of PACS and SPIRE. The disk is well resolved at 70, 100, and 160~\micron. It is resolved along its major axis at 250~\micron; at 350~\micron, the emission is point-like. The emission at 500~\micron~cannot be separated from that of a nearby background source. Our measurements are consistent with the disk being aligned with the stellar equator given the inclination of \bout70\degree~that is measured for \gamDor~based on its stellar oscillation modes \citep{Balona1996}. 

The SED of the dust emission has a shape that is too broad to arise from dust at a single temperature. There is both cool dust (observable to \Herschel) and warm dust (evident from IRS and MIPS 24 excesses). Similarly, the resolved images cannot be modeled by a single narrow dust ring, suggesting any temperature distribution within the cool dust arises from a radially broad distribution of dust and not dust at a single stellocentric radius with a distribution of grain properties.

We have modeled the resolved images, and therefore cool dust, at 70, 100, and 160~\micron~as arising from two narrow dust rings and as a single wide dust belt. The narrow rings model has two rings at large distances (70 and 190~AU) from the star and accounts for all the observed excess across mid-IR and submm wavelengths. The wide belt model has a cool, wide belt (extending from 55 to 400~AU) and accounts for far-IR and submm excesses, but not IRS and MIPS 24 excesses. Consequently, a warm inner ring must be included in this model to account for the fluxes detected at shorter wavelengths. Therefore, both models require a total of two dust components (with a total $f_d \sim 10^{-5}$) in the debris disk: two narrow rings of cool dust, or a wide belt of cool dust and a narrow ring of warm dust. As both models produce reasonably low residuals, the \Herschel~resolution is unable to distinguish which configuration best represents the \gamDor~disk.

In the context of the wide belt imaging model, the IRS and MIPS 24 excesses are attributed to a warm inner ring. Given the variability between the conclusions of different IRS observations and reductions, we do not attempt to place strong constraints on the properties of this ring, but estimate its temperature to be \bout225 \p 100~K and its fractional luminosity to be \bout $4.4 \times 10^{-6}$. We derive a blackbody radius of \bout4~AU for the warm dust ring given a 225~K temperature, although it will be larger in the case of non-blackbody grains. The fractional luminosity of this ring is within the levels that are expected for the steady state evolution of a ring at 4~AU and therefore a transient event is not needed to explain the levels of warm dust in this system. 

Planets are observed to lie between dust components in systems where both dust and planets have been observed. \gamDor~is therefore a good candidate for planet searches, particularly if planets are responsible for the lack of dust between the inner and outer dust components. The most likely region to find planets would be within 55~AU of the star, which both models support as clear of dust. Planets found beyond 55~AU would offer support to the narrow rings model as they could be responsible for the lack of dust between the two rings of cool dust.


\acknowledgments

H.B.F. and M.B. acknowledge research support from the Canadian Space Agency's Space Science Enhancement Program. 
Many thanks to Karl Stapelfeldt for the helpful exchange on comparing IRS data to photometry. 
This work is based on observations made with Herschel, a European Space Agency Cornerstone Mission with significant participation by NASA. Support for this work in part was provided by NASA through an award (No. 1353184, PI: H. M. Butner) issued by the Jet Propulsion Laboratory, California Institute of Technology under contract to NASA. 


\bibliographystyle{apj}
\bibliography{F085} 

\end{document}